\pdfoutput=1

\documentclass{llncs}

\usepackage{snapshot}
\usepackage{makeidx}

\usepackage[T1]{fontenc} \usepackage[utf8]{inputenc}
\usepackage{url}
\usepackage{graphicx}
\DeclareGraphicsExtensions{.pdf,.png,.jpg}
\graphicspath{{./img/}}
\usepackage{algorithmic}
\usepackage{flushend}
\usepackage{amsmath}
\usepackage{float}
\usepackage{amssymb}
\newcommand{\coe}{$CO_{2}e$}

\title{Energy Efficient Service Delivery in Clouds in Compliance with the Kyoto Protocol}

\author{Drazen Lucanin\inst{1,2} \and Michael Maurer\inst{1} \and Toni Mastelic\inst{1} \and Ivona Brandic\inst{1}}

\institute{Vienna Univ. of Technology, Vienna, Austria
\email{\{drazen,maurer,toni,ivona\}@infosys.tuwien.ac.at}
\and Ruder Boskovic Inst., Zagreb, Croatia}

\begin{document}

\maketitle

\begin{abstract}
Cloud computing is revolutionizing the ICT landscape by providing scalable and
efficient computing resources on demand. The ICT industry -- especially data
centers, are responsible for considerable amounts of $CO_2$ emissions and will
very soon be faced with legislative restrictions, such as the Kyoto protocol, defining caps at different organizational
levels (country, industry branch etc.) A lot has been done around
energy efficient data centers, yet there is very little work done in
defining flexible models considering $CO_2$.
In this paper we present a first attempt of modeling data centers in compliance
with the Kyoto protocol. We discuss a novel approach for trading credits for
emission reductions across data centers to comply with their constraints.
$CO_2$ caps can be integrated with Service Level Agreements and juxtaposed to other
computing commodities (e.g. computational power, storage), setting a foundation
for implementing next-generation schedulers and pricing models that support Kyoto-compliant
$CO_2$ trading schemes.
\end{abstract}

\section{Introduction}

With the global advent of cloud, grid, cluster computing and increasing needs
for large data centers to run these services, the environmental impact of
large-scale computing paradigms is becoming a global problem. The energy
produced to power the ICT industry (and data centers constitute its major part)
is responsible for 2\% of all the carbon dioxide equivalent (\coe\ -- greenhouse
gases normalized to carbon dioxide by their environmental impact) emissions
\cite{_gartner_????-1}, thus accelerating global warming \cite{_ipcc_????}.

Cloud computing facilitate users to buy computing resources from a cloud
provider and specify the exact amount of each resource (such as storage space,
number of cores etc.) that they expect through a Service Level Agreement
(SLA)\footnote{We consider the traditional business model where the desired
specifications are set in advance, as is still the case in most
infrastructure-as-a-service clouds.}. The cloud provider then honors this
agreement by providing the promised resources to avoid agreement violation
penalties (and to keep the customer satisfied to continue doing business).
However, cloud providers are usually faced with the challenge of satisfying
promised SLAs and at the same time not wasting their resources as a user very
rarely utilizes computing resources to the maximum
\cite{beauvisage_computer_2009}.

In order to fight global warming the Kyoto protocol was established by the
United Nations Framework Convention on Climate Change (UNFCCC or FCCC). The goal
is to achieve global stabilisation of greenhouse gas concentrations in the
atmosphere at a level that would prevent dangerous anthropogenic interference
with the climate system \cite{_kyoto_????}. The protocol defines control mechanisms to reduce \coe\
emissions by basically setting a market price for such emissions. Currently, flexible models
for \coe\ trading are developed at different organizational and political level
as for example at the level of a country, industry branch, or a company.
As a result, keeping track of and reducing \coe\ emissions is becoming more and
more relevant after the ratification of the Kyoto protocol. 

Energy efficiency has often been a target for research. On the one
hand, there is large body of work done in facilitating energy efficient
management of data centers as for example in \cite{_green_????-1} where current
state of formal energy efficiency control in cloud computing relies on monitoring power usage efficiency (PUE) and the
related family of metrics developed by the Green Grid Consortium. Another
example is discussed in \cite{beloglazov_adaptive_2010} where economic
incentives are presented to promote greener cloud computing policies. On the
other hand, there are several mature models for trading \coe\ obligations in
various industrial branches, as for example in the oil industry
\cite{ellerman_european_????}. Surprisingly, to the best of our knowledge there
exists no related work about the application of the Kyoto protocol to energy
efficient modeling of data centers and cloud infrastructures.

In this paper we propose a \coe-trading model for transparent scheduling of
resources in cloud computing adhering to the Kyoto protocol guidelines
\cite{ellerman_european_????}. First, we present a conceptual model for \coe\
trading compliant to the Kyoto protocol's emission trading scheme. We consider
an \emph{emission trading market (ETM)} where \emph{credits for emission
reduction (CERs)} are traded between data centers. Based on the positive or
negative \emph{CERs} of the data center, a cost is set for the environmental
impact of the energy used by applications. Thereby, a successful application scheduing decission can
be brought after considering the (i) energy costs,
(ii) \coe\ costs and (iii) SLA violation costs. Second, we propose a
\emph{wastage-penalty} model that can be used as a basis for the
implementation of Kyoto protocol-compliant scheduling and pricing models.
Finally, we discuss potential uses of the model as an optimisation heuristic in
the resource scheduler.

The main contribution of the paper are (1) definition of the conceptual
\emph{emission trading market (ETM)} for the application of Kyoto protocol for
the energy efficiency management in Clouds (2) definition of a \emph{wastage -
penalty} model for trading of \emph{credits for emission reduction (CERs)}  (3)
discussion on how the presented \emph{wastage-penalty} model can be used for
the implementation of next generation Kyoto protocol compliant energy efficient
schedulers and pricing models.

The paper is structured as follows: Section \ref{sec:related} discusses related
work. Section \ref{sec:kyoto} gives some background as to why cloud computing
might become subject to the Kyoto protocol. Section \ref{sec:model} presents our
model in a general \coe-trading cloud scenario, we then go on to define a formal
model of individual costs to find a theoretical balance and discuss the
usefulness of such a model as a scheduling heuristic. Section
\ref{sec:conclusion} concludes the paper and identifies possible future research
directions.

\section{Related Work}
\label{sec:related}

As our aim is to enable energy efficiency control in the cloud resource
scheduling domain, there are two groups of work related to ours that  deal with
the problem:
\begin{enumerate}
  \item scheduling algorithms - resource allocation techniques, from which energy cost optimisation is starting to evolve
  \item energy efficiency legislation - existing rules, regulations and best behaviour suggestions that are slowly moving from optimising the whole data center efficiency towards optimising its constituting parts
\end{enumerate}
We will examine each of these two groups separately now.

\subsection{Scheduling Algorithms}

There already exist cloud computing energy efficient scheduling solutions, such
as \cite{maurer_enacting_2011,maurer_simulating_2010} which try to minimize
energy consumption, but they lack a strict quantitative model similar to PUE
that would be convenient as a legislative control measure to express exactly how
much they alter \coe\ emission levels. From the \coe\ management perspective,
these methods work more in a best-effort manner, attempting first and foremost
to satisfy SLA constrains.

In \cite{coutinho_workflow_2011} the HGreen heuristic is proposed to schedule
batch jobs on the greenest resource first, based on prior energy efficiency
benchmarking of all the nodes, but not how to optimize a job once it is
allocated to a node - how much of its resources is it allowed to consume. A
similar multiple-node-oriented scheduling algorithm is presented in
\cite{wang_energy-efficient_2011}.

The work described in \cite{beloglazov_adaptive_2010} has the most similarities
with ours, since it also balances SLA and energy constraints and even describes
energy consumption using a similar, linear model motivated by dynamic voltage
scaling, but no consideration of \coe\ management was made inside the model.

A good overview of cloud computing and sustainability is given in
\cite{garg_environment-conscious_2011}, with explanations of where cloud
computing stands in regard to \coe\ emissions. Green policies for scheduling are
proposed that, if accepted by the user, could greatly increase the efficiency of
cloud computing and reduce \coe\ emissions. Reducing emissions is not treated as
a source of profit and a possible way to balance SLA violations, though, but
more of a general guideline for running the data center to stay below a certain
threshold.

\subsection{Energy Efficiency Legislation}

Measures of controlling energy efficiency in data centers do exist -- metrics
such as power usage efficiency (PUE) \cite{_green_????-2}, carbon usage
efficiency (CUE), water usage efficiency (WUE) \cite{_green_????} and others
have basically become the industry standards through the joint efforts of policy
makers and cloud providers gathered behind The Green Grid consortium
\cite{_green_????-1}. The problem with these metrics, though, is that they only
focus on the infrastructure efficiency -- turn as much energy as possible into
computing inside the IT equipment. Once the power gets to the IT equipment,
though, all formal energy efficiency regulation stops, making it more of a
black-box approach. For this reason, an attempt is made in our work to bring
energy efficiency control to the interior operation of clouds -- resource
scheduling.

So far, the measurement and control of even such a basic metric as PUE is not
mandatory. It is considered a best practice, though, and agencies such as the
U.S. Environmental Protection Agency (EPA) encourage data centers to measure it
by rewarding the best data centers with the Energy Star award
\cite{_energy_????}.

\section{Applying the Kyoto Protocol to Clouds}
\label{sec:kyoto}

The Kyoto protocol \cite{grubb_kyoto_1999} commits involved countries to
stabilize their greenhouse gas (GHG) emissions by adhering to the measures
developed by the United Nations Framework Convention on Climate Change (UNFCCC)
\cite{_kyoto_????-1}. These measures are commonly known as \emph{the
cap-and-trade system}. It is based on setting national emission boundaries --
caps, and establishing international emission markets for trading emission
surpluses and emission deficits. This is known as \emph{certified emission
reductions} or \emph{credits for emission reduction} (CERs). Such a trading
system rewards countries which succeeded in reaching their goal with profits
from selling CERs and forces those who did not to make up for it financially by
buying CERs. The European Union Emission Trading System (EU ETS) is an example
implementation of an emission trading market \cite{_eu_????}. Through such
markets, CERs converge towards a relatively constant market price, same as all
the other tradable goods.

Individual countries control emissions among their own large polluters
(individual companies such as power generation facilities, factories\ldots) by
distributing the available caps among them. In the current implementation,
though, emission caps are only set for entities which are responsible for more than 25
Mt\coe/year \cite{_environment_????}. This excludes individual data centers
which have a carbon footprint in the kt\coe/year range \cite{_data_????}.

It is highly possible, though, that the Kyoto protocol will expand to smaller
entities such as cloud providers to cover a larger percentage of polluters and
to increase the chance of global improvement. One such reason is that currently
energy producers take most of the weight of the protocol as they cannot pass the
responsibilities on to their clients (some of which are quite large, such as
data centers). In 2009, three companies in the EU ETS with the largest shortage
of carbon allowances were electricity producers \cite{_eu_????-1}. Another
indicator of the justification of this forecast is that some cloud providers,
such as Google already participate in emission trading markets to achieve carbon
neutrality \cite{_googles_????}.

For this reason, we hypothesize in this paper that cloud providers are indeed
part of an emission trading scheme and that \coe\ emissions have a market price.

\section{Wastage-Penalty Balance in a Kyoto-Compliant Cloud}
\label{sec:model}
In this section we present our \coe-trading model that is to be integrated with
cloud computing. We show how an economical balance can be found in it. Lastly,
we give some discussion as to how such information might be integrated into a
scheduler to make it more energy and cost efficient.

\subsection{The \coe-Trading Model}
The goal of our model is to integrate the Kyoto protocol's \coe\ trading
mechanism with the existing cloud computing service-oriented paradigm. At the
same time we want to use these two aspects of cloud computing to express an
economical balance function that can help us make better decisions in the
scheduling process.

\begin{figure}[h!]
\includegraphics[width=0.95\textwidth]{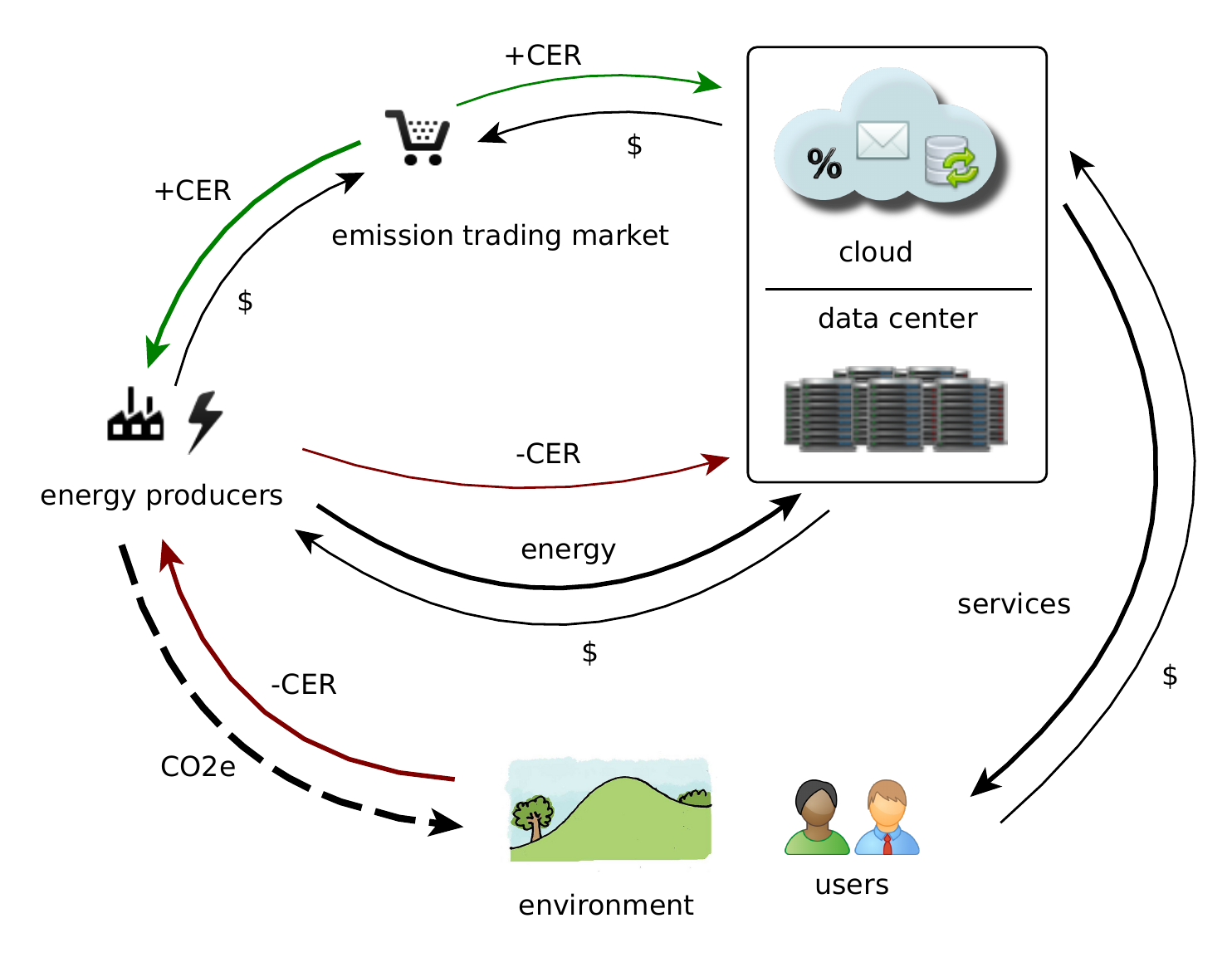}
\caption{Cloud computing integrated with the Kyoto protocol's emission trading scheme}
\label{fig:model}
\end{figure}

The model diagram in Fig. \ref{fig:model} shows the entities in our model and
their relations. A cloud offers some computing resources as services to its
users and they in turn pay the cloud provider for these services. Now, a cloud
is basically some software running on machines in a data center. To operate, a
data center uses electrical energy that is bought from an energy producer. The
energy producers are polluters as they emit \coe\ into the atmosphere. As
previously explained, to mitigate this pollution, energy producers are bound by
the Kyoto protocol to keep their \coe\ emissions bellow a certain threshold and
buy CERs for all the excess emissions from other entities that did not reach
their caps yet over the emission trading market (ETM). This is illustrated by
getting negative CERs (-CERs) for \coe\ responsibilities and having to buy the
same amount of positive CERs (+CERs) over the ETM. It does not make any real
difference for our model if an entity reaches its cap or not, as it can sell the
remaining \coe\ allowance as CERs to someone else over the ETM. Most
importantly, this means that \emph{\coe\ emissions an entity is responsible for
have a price}.

The other important thing to state in our model is that
\emph{\coe\ emission responsibilities for the energy that was bought is
transferred from the energy producer to the cloud provider}. This is shown in
Fig. \ref{fig:model} by energy producers passing some amount of -CERs to the
cloud provider along with the energy that was bought. The cloud provider then
has to buy the same amount of +CERs via the ETM (or he will be able to sell them
if he does not surpass his cap making them equally valuable).

The consequences of introducing this model are that three prices influence the
cloud provider: (1) energy cost; (2) \coe\ cost; (3) service cost. To maximize
profit, the cloud provider is motivated to decrease energy and \coe\ costs and
maximize earnings from selling his service. Since the former is achieved by
minimizing resource usage to save energy and the latter by having enough
resources to satisfy the users' needs, they are conflicting constraints. Therefore, an
economical balance is needed to find exactly how much resources to provide.

The service costs are much bigger than both of the other two combined (that is
the current market state at least, otherwise cloud providers would not operate),
so they cannot be directly compared. There are different ways a service can
be delivered, though, depending on how the cloud schedules resources. The aim of a
profit-seeking cloud provider is to deliver just enough resources to the user so
that his needs are fullfilled and that the energy wastage stays minimal. If a
user happens to be tricked out of too much of the resources initially sold to
him, a service violation occurs and the cloud provider has to pay a penalty
price. This means that we are comparing the energy wastage price with the
occasional violation penalty. This comparison is the core of our wastage-penalty
model and we will now explain how can a wastage-penalty economical balance be
calculated.

\subsection{The Wastage-Penalty Model for Resource Balancing}

As was briefly sketched in the introduction, the main idea is to push cloud
providers to follow their users' demands more closely, avoiding too much
resource over-provisioning, thus saving energy. We do this by introducing additional cost
factors that the cloud provider has to pay if he wastes too much resources  --
the energy and \coe\ costs shown in Fig. \ref{fig:model}, encouraging him to
breach the agreed service agreements and only provide what is actually needed.
Of course, the cloud provider will not breach the agreement too much, as that
could cause too many violation detections (by a user demanding what cannot be
provided at the moment) and causing penalty costs. We will now expand our model
with some formal definitions in the cloud-user interface from Fig.
\ref{fig:model} to be able to explicitly express the wastage-penalty balance in
it.

\begin{figure}
\vspace{0cm}
\includegraphics[width=1\textwidth]{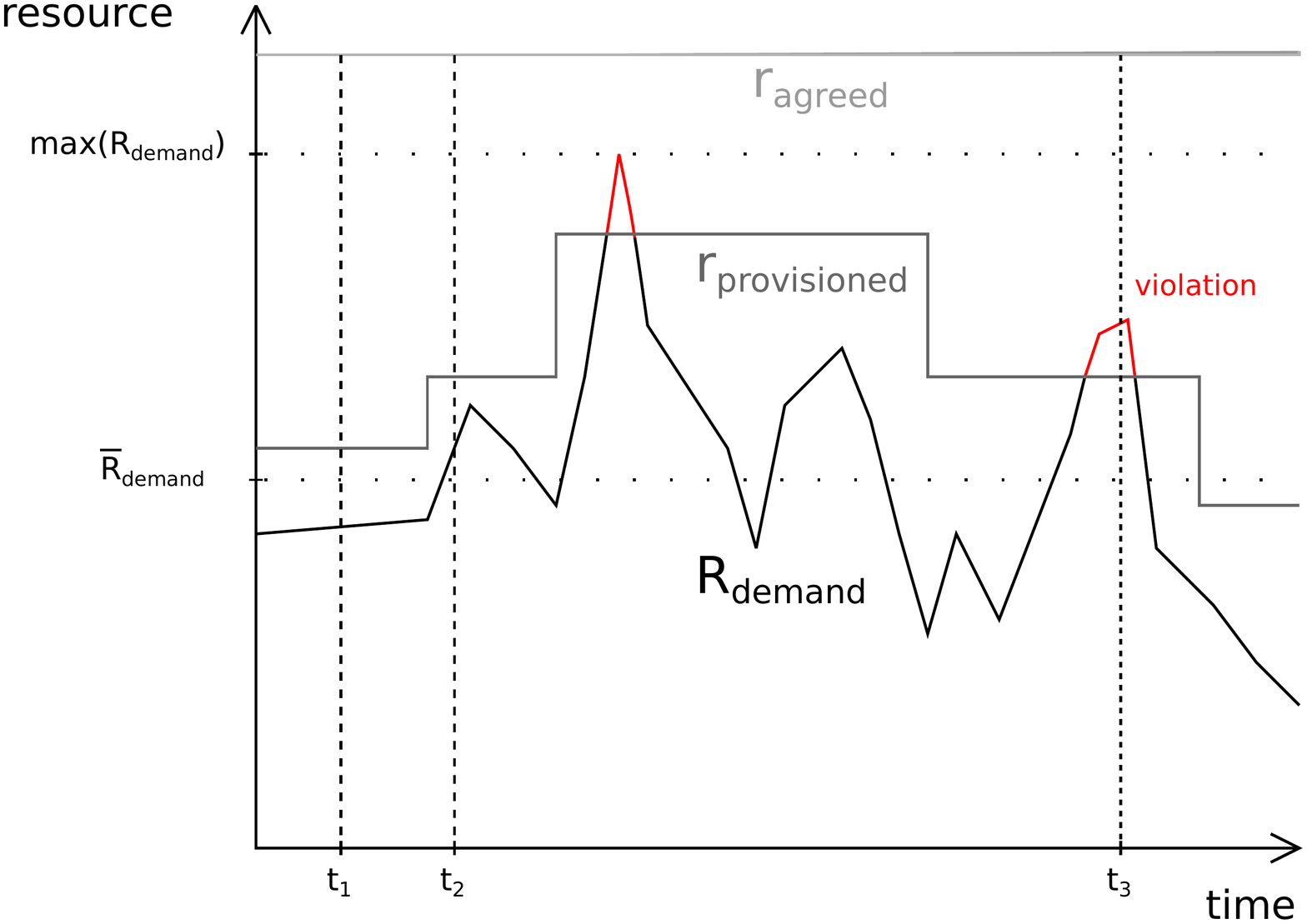}
\caption{Changes in the \emph{provisioned} and \emph{demand} resource amounts over time}
\label{fig:agreed-provisioned}
\end{figure}

We assume a situation with one cloud provider
and one cloud user. The cloud provides the user with a single, abstract resource
that constitutes its service (it can be the amount of available data storage
expressed in GB, for example). To provide a certain amount of this resource to
the user in a unit of time, a proportional amount of energy is consumed and indirectly a
proportional amount of \coe\ is emitted. An example resource scheduling scenario
is shown in Fig. \ref{fig:agreed-provisioned}. An SLA was signed that binds the
cloud provider to provide the user a constant resource amount, $r_{agreed}$. The
cloud provider was paid for this service in advance. A user uses different
resource amounts over time. At any moment the $R_{demand}$ variable is the
amount required by the user. To avoid over-provisioning the provider
does not actually provision the promised resource amount all the time, but instead adapts this
value dynamically, $r_{provisioned}$ is the resource amount allocated to the
user in a time unit. This can be seen in  Fig. \ref{fig:agreed-provisioned} as
$r_{provisioned}$ increases from $t_1$ to $t_2$ to adapt to a sudden rise in
$R_{demand}$.

As we can not know how the user's demand changes over time, we will think of
$R_{demand}$ as a random variable. To express $R_{demand}$ in an explicit way,
some statistical method would be required and research of users' behaviour
similar to that in \cite{beauvisage_computer_2009} to gather real-life data
regarding cloud computing resource demand. To stay on a high level of
abstraction, though, we assume that it conforms to some statistical distribution
and that we can calculate its mean $\overline{R}_{demand}$ and its maximum
$max(R_{demand})$. To use this solution in the real world, an appropriate
distribution should be input (or better yet -- one of several possible
distributions should be chosen at runtime that corresponds to the current user
or application profile). We know the random variable's expected value E and
variance V for typical statistical distributions and we can express
$\overline{R}_{demand}$ as the expected value $E(R)$ and $max(R_{demand})$ as
the sum of $E(R)+V(R)$ with a limited error.

\subsubsection{Wastage Costs}

Let us see how these variables can be used to model resource wastage costs. We
denote the energy price to provision the whole $r_{agreed}$ resource amount per time unit $c_{en}$ and similarly the \coe\ price $c_{co_2}$. By only using
the infrastructure to provision an amount that is estimated the user will
require, not the whole amount, we save energy that would have otherwise been
wasted and we denote this evaded wastage cost $c_{wastage}$. Since $c_{wastage}$
is a fraction of $c_{en}+c_{co_2}$, we can use a percentage $w$ to state the
percentage that is wasted:

\begin{equation}
\label{eq:cw}
c_{wastage} = w * (c_{en}+c_{co_2})
\end{equation}

We know what the extreme cases for $w$ should be -- $0\%$ for provisioning
approximately what is needed, $\overline{R}_{demand}$; and the percentage
equivalent to the ratio of the distance between $\overline{R}_{demand}$ and
$r_{agreed}$ to the total amount $r_{agreed}$ if we provision $r_{agreed}$:

\begin{equation}
w =
\begin{cases}
1-\frac{\overline{R}_{demand}}{r_{agreed}} , & \textrm{if } r_{provisioned}=r_{agreed} \\
0, & \textrm{if } r_{provisioned}=\overline{R}_{demand}
\end{cases}
\end{equation}

We model the distribution of $w$ between these extreme values using linear
interpolation: 
average resource utilization - a ratio of the average provisioned resource
amount ($\overline{r}_{provisioned}$) and the promised resource amount
($r_{promise}$):

\begin{equation}
\label{eq:w}
w = \frac{r_{provisioned} - \overline{R}_{demand}}{r_{agreed}}
\end{equation}

If we apply \ref{eq:w} to \ref{eq:cw} we get an expression for the wastage cost.:

\begin{equation}
\label{eq:wastage}
c_{wastage} = \frac{r_{provisioned} - \overline{R}_{demand}}{r_{agreed}} * (c_{en}+c_{co_2})
\end{equation}

\subsubsection{Penalty Costs}

Let us now use a similar approach to model penalty costs. If a user demands more
resources than the provider has provisioned, an SLA violation occurs. The user gets only the provisioned amount of resources in this
case and the provider has to pay the penalty cost $C_{penal}$. While $c_{en}$
and $c_{co_2}$ can be considered constant for our needs, $C_{penal}$ is a random
variable, because it depends on the user's behaviour which we can not predict
with 100\% accuracy, so we will be working with $E(C_{penal})$, its expected
value.



$E(C_{penal})$, the expected value of $C_{penal}$ can be calculated as:

\begin{equation}
\label{eq:penalty}
E(C_{penal}) = p_{viol} * c_{viol}
\end{equation}

where $c_{viol}$ is the constant cost of a single violation (although in reality
probably not all kinds of violations would be priced the same) and $p_{viol}$ is
the probability of a violation occurring. This probability can be expressed as a
function of $r_{provisioned}$, $r_{agreed}$ and $R_{demand}$, the random
variable representing the user's behaviour:

\begin{equation}
p_{viol} = f(\overline{r}_{provisioned}, r_{promise}, R_{demand})
\end{equation}

Again, same as for $c_{wastage}$, we know the extreme values we want for
$p_{viol}$. If 0 is provisioned, we have 100\% violations and if
$max(R_{demand})$ is provisioned, we have 0\% violations:

\begin{equation}
p_{viol} =
\begin{cases}
100\%, & \textrm{if } r_{provisioned}=0 \\
0\%, & \text{if } r_{provisioned}=max(R_{demand})
\end{cases}
\end{equation}

and if we assume a linear distribution in between we get an
expression for the probability of violations occuring, which is needed for
calculating the penalty costs:

\begin{equation}
\label{eq:violation}
p_{viol}= 1 - \frac{r_{provisioned}}{max(R_{demand})}
\end{equation}

\subsubsection{Combining the Two}

Now that we have identified the individual costs, we can state our goal
function. If the cloud provider provisions too much resources the $c_{wastage}$
wastage cost is too high. If on the other hand he provisions too little
resources, tightens the grip on the user too much, the $E(C_{penal})$ penalty
cost will be too high. The economical balance occurs when the penalty and
wastage costs are equal - it is profitable for the cloud provider to breach the
SLA only up to the point where penalty costs exceed wastage savings. We can
express this economical balance with the following equation:

\begin{equation}
c_{wastage} = E(C_{penal}) + \text{[customer satisfaction factor]}
\end{equation}

The \emph{[customer satisfaction factor]} could be used to model how our
promised-provisioned manipulations affect the user's happiness with the quality
of service and would be dependant of the service cost (because it might
influence if the user would be willing to pay for it again in the future). For simplicity's sake we
will say that this factor equals 0, getting:

\begin{equation}
\label{eq:costs}
c_{wastage} = E(C_{penal})
\end{equation}

Now, we can combine equations \ref{eq:wastage}, \ref{eq:costs}, \ref{eq:penalty}
and \ref{eq:violation} to get a final expression for $r_{provisioned}$:

\begin{equation}
\label{eq:equilibrium}
r_{provisioned} =\frac{max(R_{demand})*\left[\overline{R}_{demand}*(c_{en}+c_{co_2}) + r_{agreed}*c_{viol}\right]}{max(R_{demand})*(c_{en}+c_{co_2})+r_{agreed}*c_{viol}}
\end{equation}

This formula is basically \emph{the economical wastage-penalty balance}. All the
parameters it depends on are constant as long as the demand statistic stays the
same. It shows how much on average should a cloud provider breach the
promised resource amounts when provisioning resources to users so that the
statistically expected costs for SLA violation penalties do not surpass the
gains from energy savings. Vice versa also holds -- if a cloud provider
provisions more resources than this wastage-penalty balance, he pays more for
the energy wastage (energy and \coe\ price), than what he saves on SLA violations.

\subsection{Heuristics for Scheduling Optimisation with Integrated Emission
Management}

In this section we discuss a possible application of our wastage-penalty model
for the implementation of a future-generation data center. Knowing the
economical wastage-penalty balance, heuristic functions can be used to optimize
resource allocation to maximize the cloud provider's profit by integrating both
service and violation penalty prices and energy and \coe\ costs. This is useful,
because it helps in the decision-making process when there are so many
contradicting costs and constraints involved.

A heuristic might state: ``try not to provision more than $\pm x\%$ resources
than the economical wastage-penalty balance''. This heuristic could easily be
integrated into existing scheduling algorithms, such as
\cite{maurer_enacting_2011,maurer_simulating_2010} so that the cloud provider
does not stray too far away from the statistically profitable zone without
deeper knowledge about resource demand profiles. The benefits of using our
wastage-penalty model are:
\begin{itemize}
  \item a new, expanded cost model covers all of the influences from Fig. \ref{fig:model}
  \item \coe-trading schema-readiness makes it easier to take part in emission trading
  \item a Kyoto-compliant scheduler module can be adapted for use in resource
  scheduling and allocation solutions
  \item the model is valid even without Kyoto-compliance by setting the \coe\
  price $c_{co_2}$ to 0, meaning it can be used in traditional ways by weighing
  only the energy wastage costs against service violation penalties.
\end{itemize}

The wastage-penalty balance in \ref{eq:equilibrium} is a function of significant
costs and the demand profile's statistical properties:

\begin{equation}
r_{provisioned} =g(max(R_{demand}),\overline{R}_{demand},r_{agreed},c_{en},c_{co_2},c_{viol})
\end{equation}

This function enables the input of various statistical models for user or
application demand profiles ($max(R_{demand})$ and $\overline{R}_{demand}$) and
energy ($c_{en}$), \coe\ ($c_{co_2}$) and SLA violation market prices
($c_{viol}$). With different input parameters, output results such as energy
savings, environmental impact and SLA violation frequency can be compared. This
would allow cloud providers and governing decision-makers to simulate the
effects of different scenarios and measure the influence of individual
parameters, helping them choose the right strategy.




\section{Conclusion}
\label{sec:conclusion}
In this paper we presented a novel approach for Kyoto protocol-compliant
modeling of data centers. We presented a conceptual model for \coe\
trading compliant with the Kyoto protocol's emission trading scheme. We
consider an \emph{emission trading market (ETM)} where \coe\ obligations are
forwarded to data centers, involving them in the trade of credits for emission
reduction (CERs). Such measures would ensure a \coe\ equilibrium and encourage
more careful resource allocation inside data centers.

To aid decission making inside this \coe-trading system, we proposed a
\emph{wastage-penalty} model that can be used as a basis for the
implementation of Kyoto protocol-compliant scheduling and pricing models. In the
future we plan to implement prototype scheduling algorithms for the ETM
considering self-adaptable Cloud infrastructures.
\

\subsubsection{Acknowledgements} The work described in this paper was
funded by the Vienna Science and Technology Fund (WWTF) through project
ICT08-018 and by the TU Vienna funded HALEY project (Holistic Energy Efficient Management of Hybrid Clouds).

\bibliographystyle{splncs03}
\bibliography{kermit-library}

\end{document}